\documentclass[a4paper,11pt]{article}
\pdfoutput=1 

\usepackage{jcappub} 

\usepackage[T1]{fontenc} 

\title{The impact of constrained interacting dark energy on the bound-zone velocity profile}

\author[a,1]{Jounghun Lee \note{Corresponding author.}}
\author[b,c,d]{Marco Baldi }

\affiliation[a]{Department of Physics and Astronomy, Seoul National University, \\
Kwanak-ro 1, Kwanak-gu, Seoul 08826, Republic of Korea}
\affiliation[b]{Dipartimento di Fisica e Astronomia, Alma Mater Studiorum  University of Bologna, \\ 
Via Piero Gobetti 93/2, I-40129 Bologna BO, Italy}
\affiliation[c]{INAF - Osservatorio Astronomico di Bologna, \\Via Piero Gobetti 93/3, I-40129 Bologna BO, Italy}
\affiliation[d]{INFN - Istituto Nazionale di Fisica Nucleare, Sezione di Bologna, \\ Viale Berti Pichat 6/2, I-40127 Bologna BO, Italy}

\emailAdd{cosmos.hun@gmail.com}
\emailAdd{marco.baldi5@unibo.it}

\abstract{We numerically study the effects of constrained interacting dark energy (CIDER) on the bound-zone velocity profiles around massive dark matter halos.
Analyzing the CIDER simulations performed by ref.~\cite{cider} for three different cases of dark sector coupling ($\beta=0.03$, $0.05$ and $0.08$)
as well as for the standard $\Lambda$CDM cosmology ($\beta=0$), we determine the mean peculiar velocity profiles in the bound zones around the friends-of-friends 
halos with masses larger than $M_{\rm cut}=3\times 10^{13}\,h^{-1}M_{\odot}$ at three redshifts, $z=0$, $0.5$ and $1$. It is found that the universal power-law formula 
proposed by ref.~\cite{fal-etal14} originally for the $\Lambda$CDM case still describes well the bound-zone velocity profiles, $V(r)$, even in the CIDER models. 
The slope of $V(r)$, turns out to be significantly affected by the CIDER,  progressively decreasing as $\beta$ increases.  
Meanwhile, the amplitude of $V(r)$ exhibits little dependence on $\beta$, which is ascribed to  the identical Hubble parameters shared by the $\Lambda$CDM and 
CIDER models in the entire redshift range. Our results imply that the bound-zone velocity slope can break a degeneracy even between the $\Lambda$CDM and 
CIDER models with $\beta\le 0.03$, which the standard cosmological diagnostics fail to distinguish.  We devise a simple analytic formula for the bound-zone slope 
as a function of $\beta$, and prove its validity at all of the three redshifts. It is concluded that the slope of the mean bound-zone peculiar velocity profile 
should be in principle a powerful probe of dark sector interaction. }
\begin{document}
\maketitle
\flushbottom

\section{Introduction}\label{sec:intro}

It has been well known that as the peculiar velocity field contains an independent dynamical information on the background cosmology~\cite{fel03}, 
it is most suitable for testing degenerate non-standard cosmologies like dynamical dark energy and modified gravity scenarios~\cite{zu-etal14,hel-etal14,hel-etal16,LB22}. 
In the seminal work of ref.~\cite{fal-etal14} it was numerically proven that in the bound-zones around massive dark matter (DM) halos where the gravitational attraction is greatly diluted 
by the cosmic acceleration in the late-universe, the peculiar velocity behaves like a linear variable and thus its radial profile can be analytically tractable. 
Here, a bound-zone of a massive halo with virial radius, $r_{\rm v}$, refers to the distance range, ($3-8$)$r_{\rm v}$ at which its neighbor lower-mass halos 
recede from it at lower speeds than the global Hubble flow due to the halo's gravitational attraction~\citep{fal-etal14}. It is located in-between the infall ($< 3r_{\rm v}$) and Hubble 
($> 8r_{\rm v}$) zones where the recession speeds become essentially zero and equal to the global value, respectively \citep{zu-etal14,lee16}.  

Multiple numerical works have shown that the bound-zone velocity profile is well approximated by the following universal formula with two adjustable parameters $A$ and 
$n$~\cite{pra-etal06,cue-etal08,fal-etal14,lee16,LB22}:
\begin{equation}
\label{eqn:vr}
\frac{V(r)}{V_{\rm v}} = A\left(\frac{r}{r_{\rm v}}\right)^{-n}\,  ,
\end{equation}
where $V(r)$ is the mean radial component of the peculiar velocity field at a bound-zone distance $r$ from a massive halo with mass $M$, while $V_{\rm v}$ is the 
radial component of virial velocity defined as $V^{2}_{\rm v}\equiv GM/r_{\rm v}$. The universality of eq.~(\ref{eqn:vr}) is manifested by the weak dependences of $n$ and $A_{\rm }$ on 
$M$~\citep{fal-etal14} as well as on the initial conditions of the concordance model~\cite{lee16} where the gravitational law is described by the general relativity (GR),  dark energy (DE) 
responsible for the cosmic acceleration is the cosmological constant $\Lambda$, and the most dominant source of gravity is the cold dark matter (CDM). 

This universal form of the bound-zone velocity profile is believed to be established  by a delicate counter-balance between the gravity and accelerating spacetime~\cite{lee16}. 
Its amplitude, $A$, measures the overall strength of gravitational influence at the boundary between the infall and bound zones where the global Hubble flow begins to be dominant. 
Meanwhile,  its slope, $n$,  measures how rapidly the gravitational influence wanes in the bound zone due to the cosmic acceleration. 
Thus, it is naturally expected that any deviation of the background from the $\Lambda$CDM cosmology would affect the shape of $V(r)$.
Ref.~\cite{LB22} explored if and how the shape of $V(r)$ would change in the presence of modified gravity (MG) and massive neutrinos ($\nu$). According their results, 
a substantial difference exists in the best-fit values of $A$ and $n$ between the $\Lambda$CDM cosmology and a certain MG+$\nu$ model,  which cannot be discriminated by 
the conventional cosmological diagnostics~\cite{bal14}. 

The shape of the bound-zone velocity profile, $V(r)$, will be particularly useful to test those non-standard cosmologies which affect only one of the two parameters, $n$ and $A$ in 
eq.~(\ref{eqn:vr})  and thus suffer from no extra degeneracy between the two parameters.  For example, if a non-standard cosmology differs from the $\Lambda$CDM in 
its growth history but not in the expansion counterpart, then, it would leave the amplitude, $A$, of $V(r)$ intact, modifying only its slope, $n$,  since $A$ depends most sensitively on the 
value of the Hubble parameter at a given epoch~\cite{fal-etal14,lee16,LB22}.  
The recently proposed {\it constrained interacting dark energy} (CIDER) models~\cite{bar-etal19} are such non-standard models as could be powerfully tested by the slope of $V(r)$.  

The CIDER is a special type of interacting scalar field DE, $\phi$, coupled with the CDM through energy-momentum 
exchange~\cite{wet95, ame00, FP04, HW06,ame-etal08,sim10, bal11a, bal12,ame-etal14,sko-etal15}.  
In the classical interacting DE models,  the DE-CDM coupling have two prolonged effects on the dark sector. 
First, it drives the dynamical evolution of the DE density and its equation of stage. Second, it promotes the density growths by exerting an attractive fifth force on the CDM particles.  
By tracing the cosmic expansion and growth histories, it is in principle possible to constrain the strength of DE-CDM coupling, which enters as a coefficient in the scalar field DE 
potential, $U(\phi)$~\cite{wet95,ame00,FP04,sim10}. 
A critical difference between the CIDER and the classical interacting DE lies in the fact that the shape of $U(\phi)$ is specified {\it a priori} for the latter but not for the former, 
whose dynamical evolution instead determines $U(\phi)$ under the condition of identical Hubble parameter to the $\Lambda$CDM case in the whole redshift range. 
In other words, the $\Lambda$CDM and CIDER models can be differentiated from each other only through their difference in the growth histories, as the two share the same expansion history.  

The CIDER scenario has recently garnered probing attention as it turned out to be capable of alleviating the $\sigma_{8}$ tension that the local estimates of the linear density power spectrum 
amplitude, $\sigma_{8}$, is significantly lower than the best-fit value from the cosmic microwave background radiation temperature spectrum  (see ref.~\cite{s8_tension} and references therein 
for a detailed review of the current status on  the $\sigma_{8}$ tension).  
In contrast to the majority of other interacting DE and MG effects,  a lower $\sigma_{8}$ is predicted by the CIDER models in which the dark sector coupling has an effect of suppressing 
the density growths, the degree of which increases with the coupling strength~\cite{bar-etal19}. 
Yet, no large difference in the growth history can be accommodated by the current observational results between the $\Lambda$CDM and a viable CIDER model. 
In other words, a CIDER model is viable only provided that it is almost degenerate with the $\Lambda$CDM, having very weak dark sector couplings. 

Given this distinct features of the CIDER models, we speculate that the CIDER might affect only the slope, $n$, of the mean bound-zone velocity profile with amplitude $A$ intact, 
and thus that $V(r)$ might be able to break the degeneracy between the $\Lambda$CDM and a viable CIDER model.
Here, we will test our speculation against a large-volume simulation performed by ref.~\cite{cider} for three different CIDER models. The organization of this paper as follows.
In section~\ref{sec:review}, we will briefly review the datasets from the CIDER simulations. 
In section~\ref{sec:slope}, we will explain how to measure the mean bound-zone velocity profiles from simulation data and present what effects the CIDER has on the bound-zone 
velocity slopes. In section~\ref{sec:model}, we will present a simple fitting formula for the dependence of bound-zone velocity slope on the dark sector coupling. 
In section~\ref{sec:con}, we will summarize the results and draw a conclusion.

\section{Bound-zone velocity profiles in the CIDER models}

\subsection{The CIDER simulations: a brief review}\label{sec:review}
\begin{figure}[tbp]
\centering 
\includegraphics[width=0.85\textwidth=0 380 0 200]{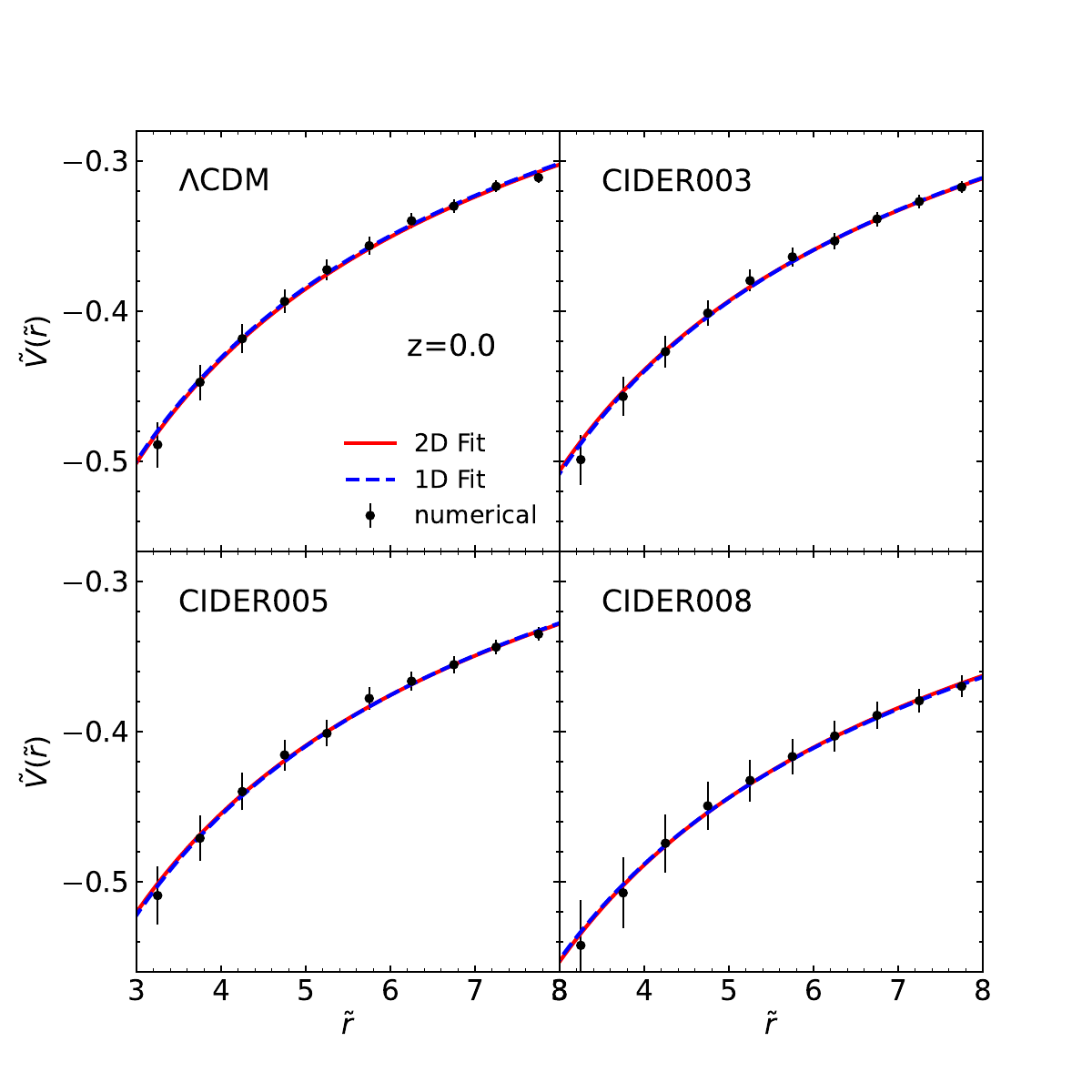}
\caption{\label{fig:v0} Numerically obtained bound-zone velocity profiles (black filled circles) compared with the analytic 
single parameter (red solid line) and double parameter (blue dashed line) formulae for three different CIDER models as well as 
for the standard $\Lambda$CDM cosmology at $z=0$.}
\end{figure}
\begin{figure}[tbp]
\centering 
\includegraphics[width=0.85\textwidth=0 380 0 200]{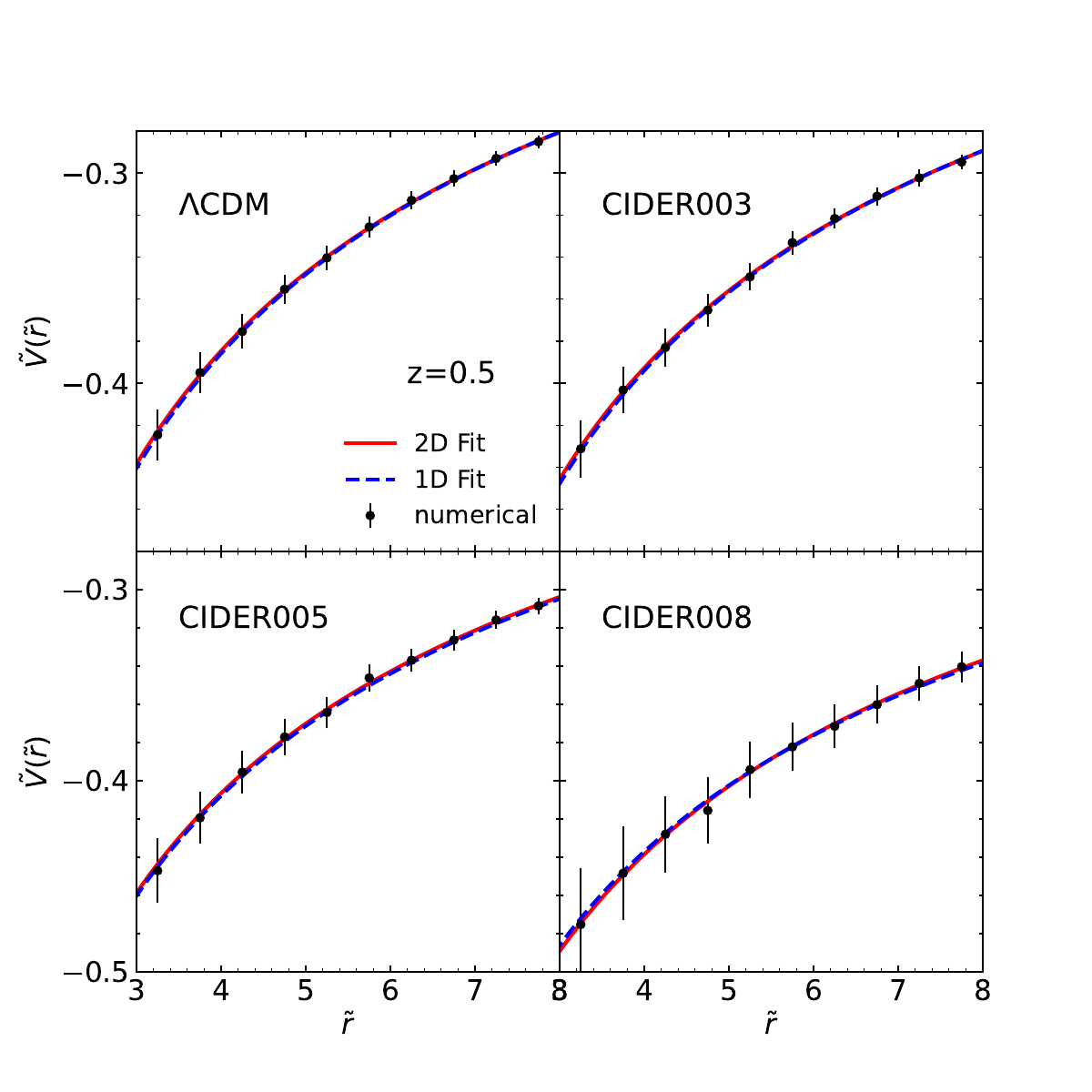}
\caption{\label{fig:v0.5} Same as figure~\ref{fig:v0} but at $z=0.5$. }
\end{figure}
\begin{figure}[tbp]
\centering 
\includegraphics[height=410 pt,width=395 pt]{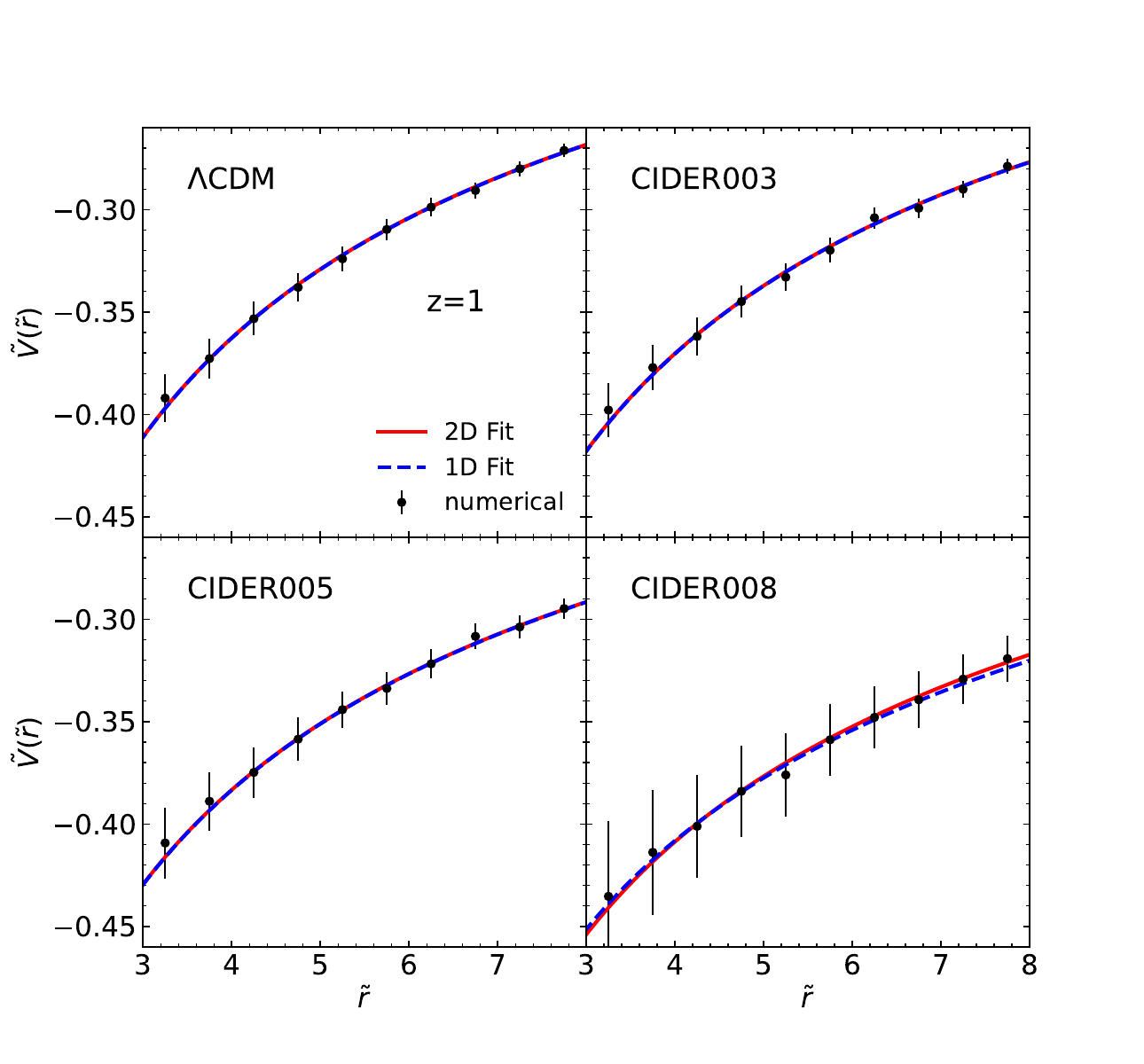}
\caption{\label{fig:v1} Same as figure~\ref{fig:v0} but at $z=1$.}
\end{figure}
\begin{figure}[tbp]
\centering 
\includegraphics[width=0.85\textwidth=0 380 0 200]{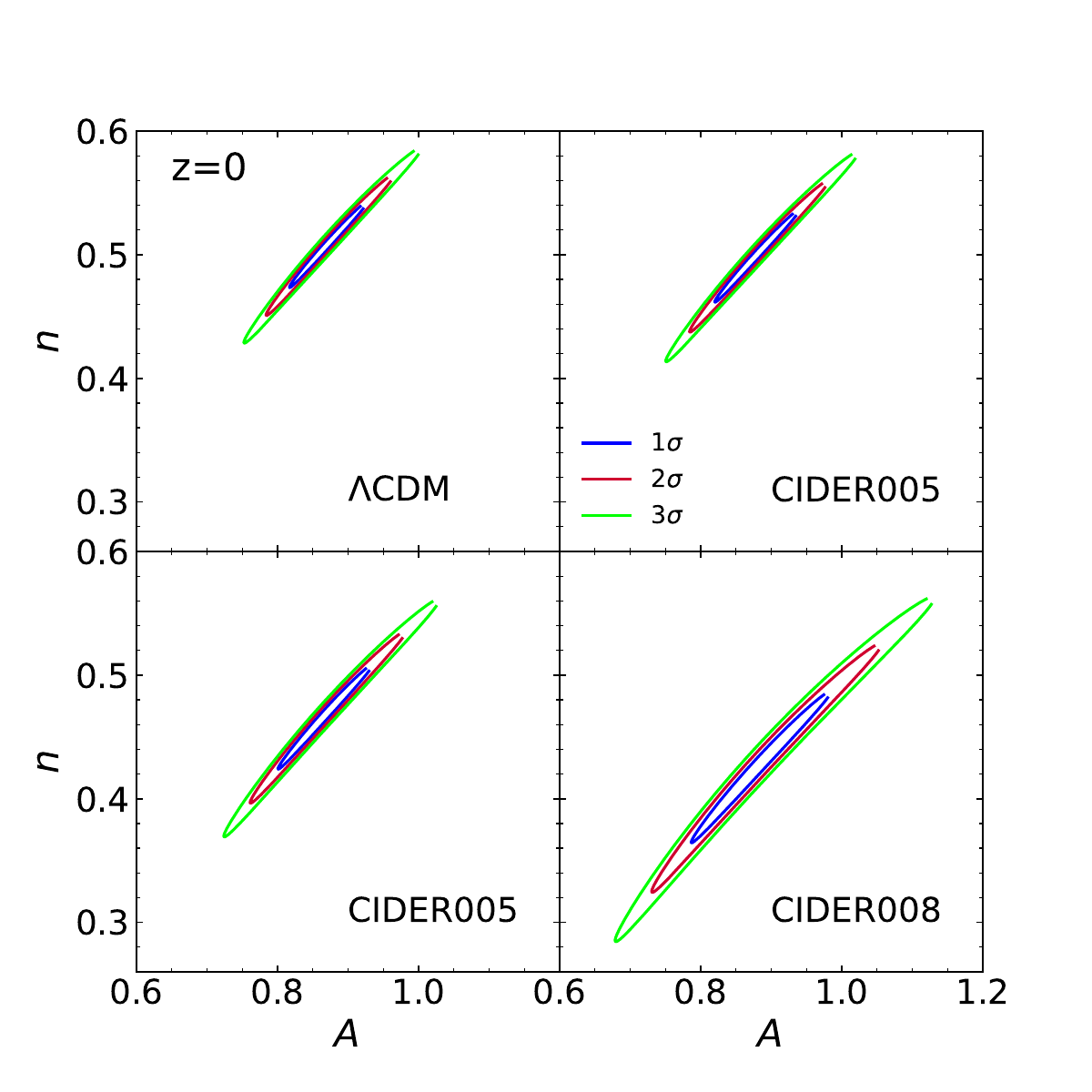}
\caption{\label{fig:cont0} $68\%$, $95\%$ and $99\%$ contours of $\chi^{2}(A,n)$ between the numerical and analytical 
results of the bound-zone velocity profiles in the 2D plane spanned by $A$ and $n$ that denote two 
adjustable parameters of the analytical formula given in eq.~(\ref{eqn:vr}) for the four cosmologies at $z=0$.}
\end{figure}
\begin{figure}[tbp]
\centering 
\includegraphics[width=0.85\textwidth=0 380 0 200]{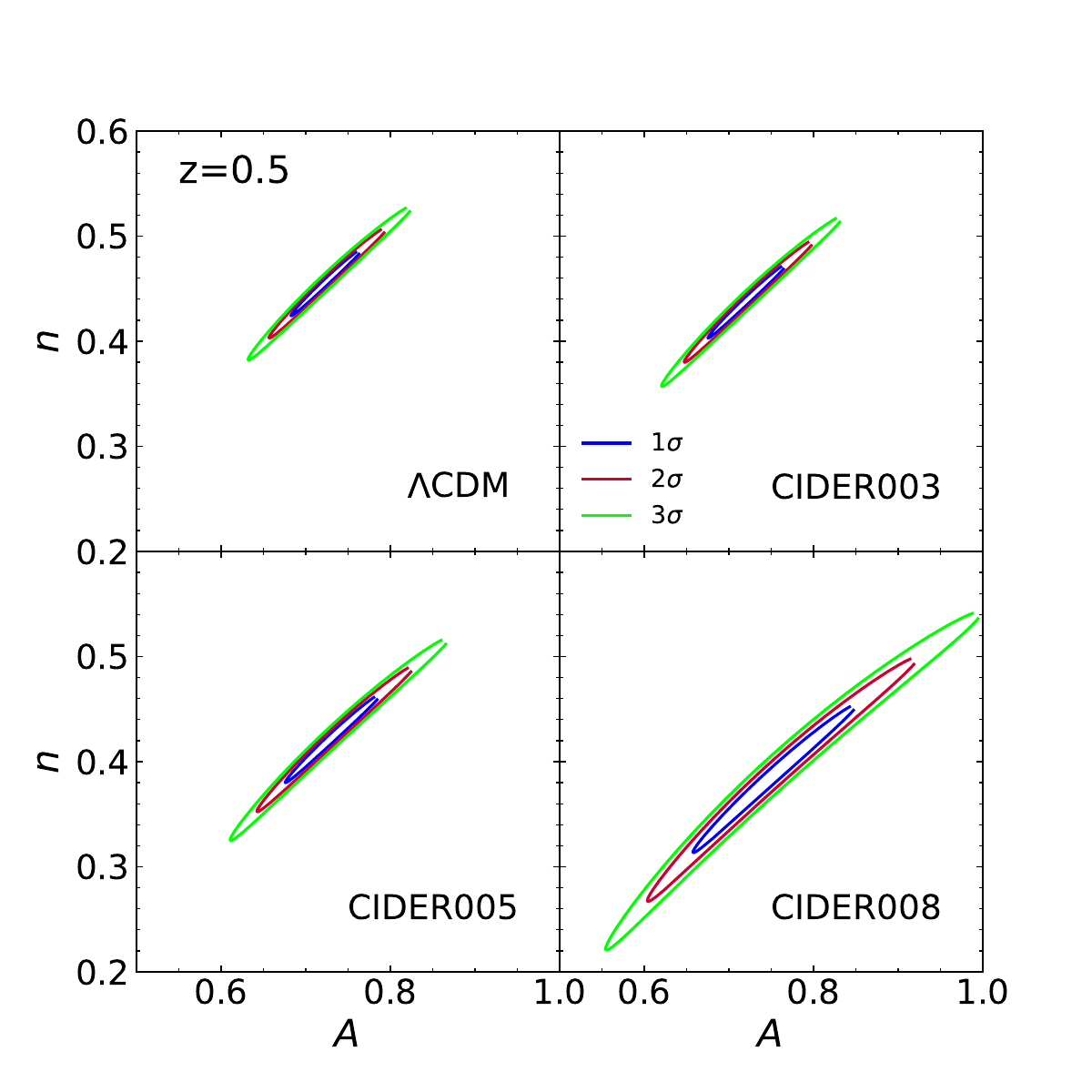}
\caption{\label{fig:cont0.5} Same as figure~\ref{fig:cont0} but at $z=0.5$.}
\end{figure}
\begin{figure}[tbp]
\centering 
\includegraphics[height=410 pt,width=395 pt]{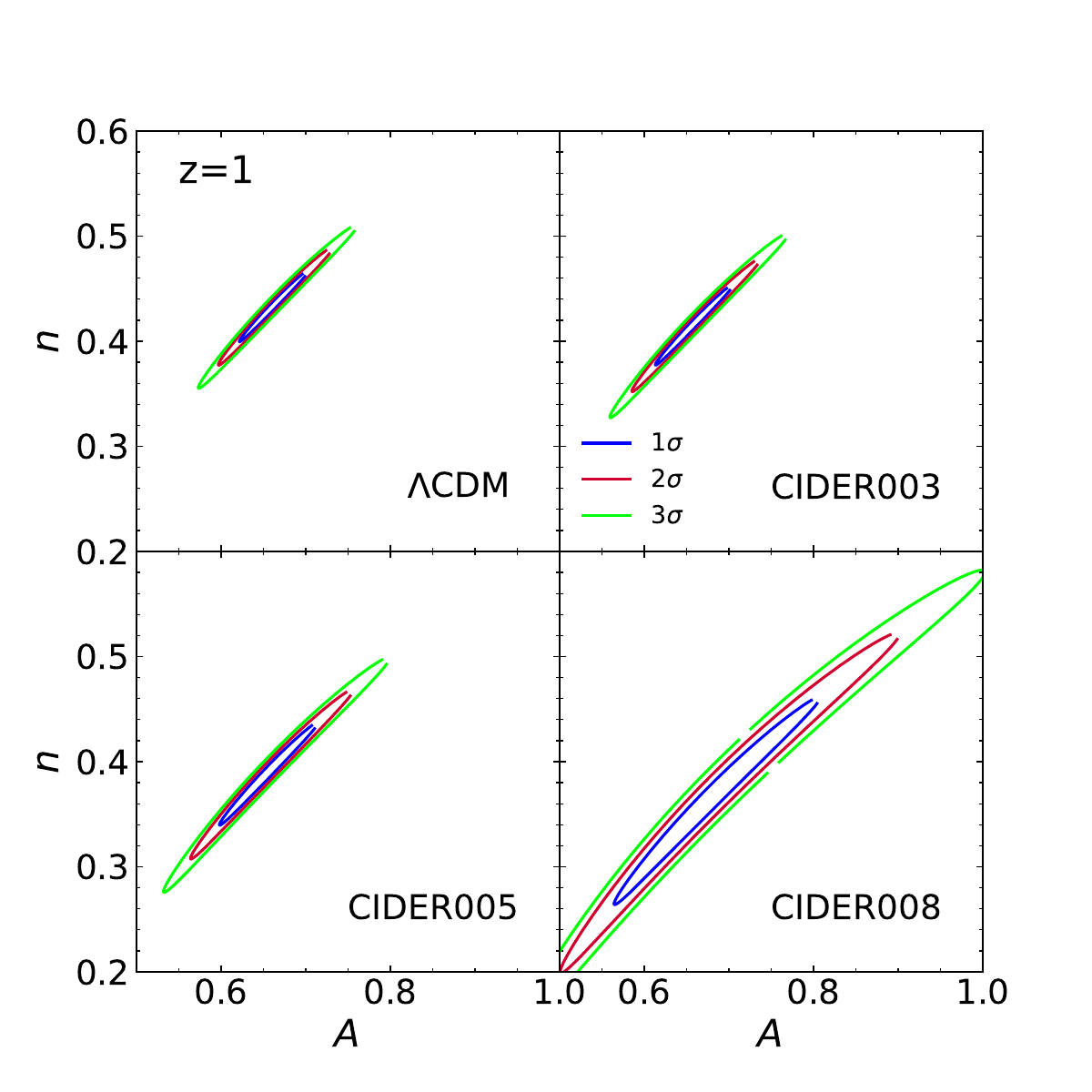}
\caption{\label{fig:cont1} Same as figure~\ref{fig:cont0} but at $z=1$.}
\end{figure}

To numerically explore the effect of CIDER on the mean bound-zone velocity profile, we utilize the halo catalogs extracted from the CIDER simulation~\cite{cider}
that belongs to a second generation of the CoDECS~\footnote{It stands for COupled Dark Energy Cosmological Simulations} project~\cite{bal12}, a suite of $N$-body simulations 
for interacting DE models, called the CoDECS 2 project. 
As most of the classical interacting DE models considered in the original project had gone out of favor with current observational data~\cite{pet13,planck15,GV20}, 
the CoDEC 2 has been launched to venture various new interacting DE scenarios including the CIDER.   The unique constraint that the Hubble parameter, $H(z)$, follows the identical redshift evolution 
to that of the $\Lambda$CDM case characterizes the CIDER models whose potential, $U(\phi)$,  can be derived from this constraint under the assumption that the DE-CDM coupling 
does not vary with time~\cite{bar-etal19}.  Unlike in the majority of the interacting DE and MG models, the matter density field in the CIDER models have suppressed powers~\cite{bar-etal19}, 
the degree of which is determined by the value of its dark sector coupling constant, $\beta$~\cite{cider}. 

The CIDER simulations were performed by ref.~\cite{cider} on a periodic box of side length $1\,h^{-1}$Gpc for four different cosmologies: the $\Lambda$CDM, 
CIDER003, CIDER005, and CIDER008 corresponding to $\beta=0,\ 0.03,\ 0.05$ and  $0,08$, respectively.  According to the results of ref.~\cite{cider}, 
the CIDER models with $\beta\le 0.03$ could not be discriminated from the $\Lambda$CDM by the standard linear diagnostics such as linear density power spectrum 
and linear growth fact, while the extreme CIDER008 model can be safely ruled out by the current observational data~\cite{planck15}.  
Except for the density power spectrum amplitude ($\sigma_{8}$),  these four models share the identical initial conditions 
of $h=0.677$, $\Omega_{\rm de}=0.689$, $\Omega_{m}=0.262$, $\Omega_{b}=0.049$, and $n_{s}=0.9665$, with dimensionless Hubble parameter ($h$), 
DE density parameter ($\Omega_{\rm de}$),  matter density parameter ($\Omega_{m}$) , baryon density parameter ($\Omega_{b}$) and spectral index ($n_{s}$). 
The values of $\sigma_{8}$ for these four models are listed in the third columns of table~\ref{tab:model}. 

Starting at redshift $z_{i}\sim 99$ with these initial conditions, the CIDER simulations tracked down the dynamical evolution of $1024$ cDM particles and 
equal number of baryonic particles as well, with the help of the C-Gadget code~\cite{bal-etal10}, an adapted version of the Gadget 3 code~\cite{gadget} for the 
computation of dark sector interaction. With the bound halos identified from the CIDER snapshots via the standard friends-of-friends (FoF) algorithm with a linkage length 
parameter of $0.2$, ref.~\cite{cider} evaluated the halo mass functions, bias functions, density profiles and concentration-mass relations for each model. It was found 
that even these nonlinear statistics could not tell apart the $\Lambda$CDM from the CIDER models if $\beta\le 0.03$.
For a comprehensive description of the CIDER simulations, we refer the readers to ref.~\cite{cider}.

\subsection{The effect of dark sector coupling on the slope of $V(r)$}\label{sec:slope}

For each model, we determine the mean bound-zone velocity profiles averaged over the massive FoF halos at three different redshifts, $z=0,\ 0.5$ and $1$, by taking the following 
steps~\cite{LB22}:
\begin{enumerate}
\item 
Select those FoF halos with total mass $M\ge 3\times 10^{13}\,h^{-1}\,M_{\odot}$. Determine the virial radii of the selected halos as
$r_{\rm v}\equiv \left(3M/[4\pi\,\Delta\rho_{\rm crit}(z)]\right)^{1/3}$ where critical density of the universe $\rho_{\rm crit}(z)$ and halo overdensity threshold 
$\Delta \equiv 100$~\cite{fal-etal14}. 
\item
Using each selected halo as a target, search for its bound-zone neighbor halos located at radial separation distances, $r\in (3-8)r_{\rm v}$. 
If any of the bound-zone neighbors turn out to be more massive than the target, then exclude the selected halo from the target list. 
From here on, the neighbor lower-mass halos located in the bound-zone of a target halo will be referred to as the bound-zone neighbors.
\item
For each target halo, compute the relative radial velocities of its bound-zone neighbors and rescale them as $\tilde{V}\equiv V/V_{\rm v}$. 
Also, rescaling their separation distances as $\tilde{r}\equiv r/r_{\rm v}$, divide the range of $3\le \tilde{r}\le 8$ into ten short intervals of equal length, $\Delta\tilde{r}=0.5$.
\item 
Find the bound-zone neighbors of all target halos that fall in each $\tilde{r}$ bin. Compute the average of $\tilde{V}$ and its one standard deviation error $\sigma(\tilde{V})$   
over the stacked bound-zone neighbors at each $\tilde{r}$ bin. 
\item
Compare the numerically obtained $\tilde{V}(\tilde{r})$ with the formula, $A\tilde{r}^{-n}$ over the whole range of $\tilde{r}$ to determine the best-fit values of $A$ and $n$ 
via the minimization of $\chi^{2}(A,n)$:
\begin{equation}
\chi^{2}(A,n)\equiv \frac{1}{(N_{\rm d.o.f})}\sum_{i=1}^{N_{\rm bin}}\frac{\left[\log\tilde{V}(\tilde{r}_{i})-\log A+n\log\tilde{r}_{i}\right]^{2}}{\sigma^{2}({\tilde{V}})}\, ,
\end{equation}
\end{enumerate}
where $N_{\rm bin}=10$ is the number of $\tilde{r}$ bins, and $N_{\rm d.o.f.}$ is the degree of freedom, equal to $N_{\rm bin}-2$ as the analytic formula has 
two free parameters.
Figures~\ref{fig:v0}-\ref{fig:v1} plot the mean bound-zone velocity profiles (filled black circles) obtained via the above steps for four cosmological models, 
at $z=0,\ 0.5$ and $1$, respectively.  The universal formula, eq.~(\ref{eqn:vr}), with the best-fit values of $A$ and $n$ is also shown (red solid line) 
for comparison in each panel, confirming that this formula for the bound-zone velocity profile is indeed valid even in the CIDER models. Note also 
that the errors in $\tilde{V}(\tilde{r})$ appear to increase as $\beta$ increases, which should be attributed to the smaller numbers of the selected 
massive halos in the CIDER models with higher $\beta$ due to their more severely suppressed density powers (see table~\ref{tab:model}).  

Figures~\ref{fig:cont0}-\ref{fig:cont1} plot the $68\%,\ 95\%$ and $99\%$ contours of $\chi^{2}(A,n)$ (blue, red and green curves) in the two-dimensional 
space spanned by $A$ and $n$ for each case.  Note again that as $\beta$ increases, the sizes of the contours become larger due to the larger errors in $\tilde{V}$ 
caused by the smaller numbers of the selected massive halos, as listed in table~\ref{tab:model}. 
As can be seen, while the slope parameter, $n$, exhibits a sensitive variation with $\beta$ at each redshift, 
the amplitude,$A$, appears to be almost independent of $\beta$.  The bottom panel of figure~\ref{fig:nv_amp} plots the best-fit values of $A$ versus the scale factor, 
$a\equiv 1/(1+z)$, from the four cosmologies. For this plot, the scale factor values of $A(a)$ are slightly offset for visibility purposes.
 As $\beta$ increases, the best-fit slope, $n$, becomes lower, which indicates that the stronger dark sector coupling causes the gravitational influence in the bound zones 
to wane less rapidly. 

Noting that the dark sector coupling does not affect the amplitude of the bound-zone velocity profiles for the case of CIDER models, 
we regard eq.~(\ref{eqn:vr}) as a single parameter formula and repeat the procedure of fitting $V(r)$ to eq.~(\ref{eqn:vr}) by adjusting only $n$. 
The amplitude, $A$, is fixed  at the best-fit value of the $\Lambda$CDM case at each redshift, and the degree of freedom is accordingly set at $N_{\rm bin}-1$ 
in this procedure. In figures~\ref{fig:v0}-\ref{fig:v1}, the results of this one single-parameter fitting  is shown (blue dashed lines) for each case.  
As can be seen, despite that $A$ is fixed, the overall agreements between the numerical and analytical results are excellent for every case. 
The top-panel of figure~\ref{fig:nv_amp} plots the resulting best-fit values of $n$ versus scale factors, revealing that the four models 
yield significantly different values of $n$ at each epoch.  Even between the $\Lambda$CDM and CIDER003 models, the difference in the best-fit value of $n$ 
is found to be as significant as $\ge 4\sigma$ at each redshift, which implies that the bound-zone velocity slope could in principle break the degeneracy 
between the $\Lambda$CDM and CIDER models with $\beta\le 0.0.3$. 

\subsection{Modeling the $\beta$-dependence of bound-zone velocity slope}\label{sec:model}
\begin{figure}[tbp]
\centering 
\includegraphics[width=0.85\textwidth=0 380 0 200]{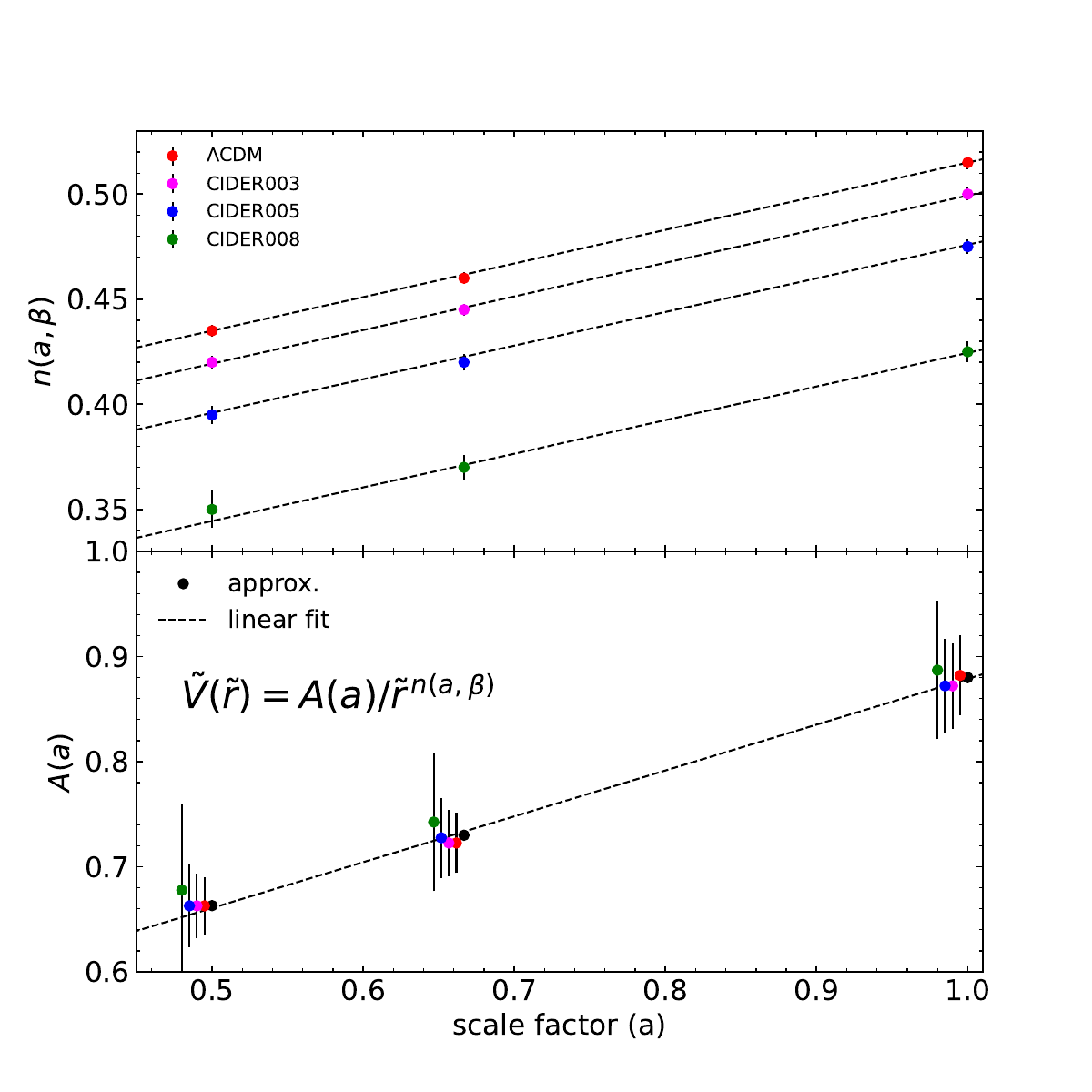}
\caption{\label{fig:nv_amp} Best-fit values of the power-law index (top panel) and amplitude (bottom panel) versus scale factor, 
compared with the linear fits (black dashed lines) for the four cosmologies. To make the data points visible, we place the 
off-set.}
\end{figure}
\begin{figure}[tbp]
\centering 
\includegraphics[width=0.85\textwidth=0 380 0 200]{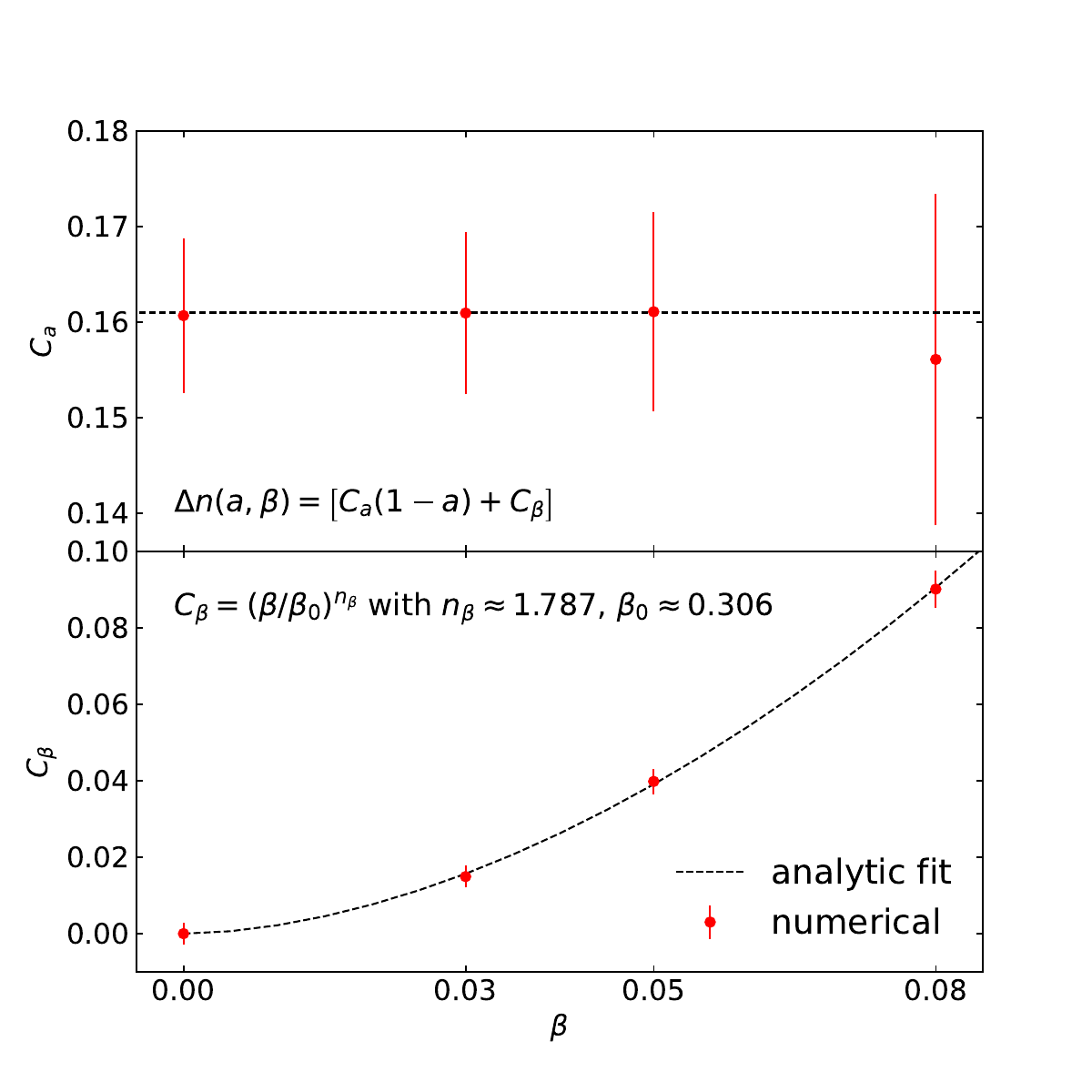}
\caption{\label{fig:coeff} $\beta$ dependence of two linear-fit constants (red filled circles) to the power-law index of the bound-zone 
velocity profile, compared with the approximate formula (black dashed lines)  where $\beta$ denotes the dark sector coupling constant.}
\end{figure}

Now that the best-fit slope $n$ is found to vary almost linearly with the scale factor $a$ for each case (figure~\ref{fig:nv_amp}), we model the differences in the bound-zone 
slope between the $\Lambda$CDM and a CIDER model at a given epoch, $n(a,\beta)$, as the following linear function: 
$\Delta n(a,\beta)$:
\begin{equation}
\label{eqn:ca_cbeta}
\Delta n(a,\beta)\equiv C_{a}(1-a) + C_{\beta} \, ,
\end{equation}
where $C_{a}$ represents the slope of this linear fit, while $C_{b}$ is the value of $\Delta n$ at the present epoch ($a=1$). 
Through the $\chi^{2}(C_{a},C_{\beta})$-minimization, we also determine the best-fit values of $C_{a}$ and $C_{\beta}$, which are shown 
in the top and bottom panels of figure~\ref{fig:coeff}, respectively.  As can be seen, $C_{a}$ shows little variation with $\beta$, while $C_{\beta}$ appears 
to be proportional to some power of $\beta$. 

We model the observed $\beta$ dependence of $C_{\beta}$ as a power-law function: 
\begin{equation}
\label{eqn:coeff}
C_{\beta} \equiv \left(\frac{\beta}{\beta_{0}}\right)^{n_{\beta}}\, . 
\end{equation}
Employing the $\chi^{2}$ statistics again, we determine the best-fit values of $\beta_{0}$ and $n_{\beta}$ to be $0.301\pm 0.079$ and $1.781\pm 0.156$, respectively, 
where the associated errors are the marginalized ones.  The bottom panel of figure~\ref{fig:coeff} shows this best-fit power-law function (dashed line)  revealing that 
$\Delta n$ at the present epoch is indeed well described by our power-law function, eq.~(\ref{eqn:coeff}). 

\begin{table}[tbp]
\centering
\begin{tabular}{ccccccc}
\hline
\hline
\rule{0pt}{4ex} 
Model & $\beta$ & $\sigma_{8}$ & $N_{h}(z=0)$ & $N_{h}(z=0.5)$ & $N_{h}(z=1)$  \medskip\\
\hline
\rule{0pt}{4ex}    
$\Lambda$CDM & $0.00$ & $0.788$ & $151142$ & $92139$ & $43462$ \medskip \\
CIDER003 & $0.03$  & $0.763$ & $135339$ & $78806$ & $35369$   \medskip \\
CIDER005 & $0.05$ & $0.723$ &  $109621$ & $58922$ & $23777$  \medskip \\
CIDER008 & $0.08$ & $0.642$ &  $60748$ & $26093$ & $8126$  \medskip \\
\hline
\end{tabular}
\caption{\label{tab:model} Model, dark sector coupling constant ($\beta$)  density power spectrum amplitude ($\sigma_{8}$)  and 
numbers of the selected massive halos  ($N_{h}$) with $M\ge 3\times 10^{13}\,h^{-1}\,M_{\odot}$ at $z=0,\ 0.5$ and $1$. }
\end{table}

 \section{Summary and conclusion}\label{sec:con}

We have numerically examined the usefulness of the mean bound-zone velocity profiles as a test of the constrained interacting dark energy (CIDER) models, 
which are believed to be capable of alleviating the $\sigma_{8}$ tension.
For this investigation, we have relied on the CIDER simulations~\cite{cider} performed for three CIDER models, which differ from one another in the 
value of dark sector coupling constant ($\beta$) but share the same expansion history and key cosmological parameters other than $\sigma_{8}$ 
as the $\Lambda$CDM case~\cite{bar-etal19}.  

Analyzing the FoF halos from the CIDER simulations, we have numerically determined the mean bound-zone velocity profiles,  $V(r)$, of massive halos 
with $M\ge 3\times 10^{13}\,h^{-1}\,M_{\odot}$ at $z=0,0.5$ and $1$ for the $\Lambda$CDM and three CIDER models. 
It has been found that the universal power-law formula given by ref.~\cite{fal-etal14} excellently describes $V(r)$ even in the CIDER models and that the CIDER 
affects only the slope of $V(r)$ with its amplitude intact. This $\beta$-independence of the bound-zone velocity amplitude has been interpreted as a consequence 
of the same expansion history. 

Noting that the bound-zone velocity slope has a progressively lower value as $\beta$ increases,  we have modeled $V(r)$ by a single-parameter power-law formula 
with constant amplitude, and found excellent agreements with the numerically obtained $V(r)$.   Showing the high statistical significances of the differences in the bound-zone 
velocity slope between the $\Lambda$CDM and CIDER models, we have proven that the single parameter, i.e., the bound-zone velocity slope, can in principle break the degeneracy 
between the $\Lambda$CDM and those CIDER models with very low dark sector coupling constant, $\beta\le 0.03$, which cannot be distinguished by the standard cosmological 
diagnostics~\cite{cider}. 
It has also been shown that the differences in the bound-zone velocity slope  between the $\Lambda$CDM and CIDER models behave like some powers of $\beta$  at the present epoch, 
and evolve almost linearly with the scale factor, $a$.  Given this result, we have devised a simple power-law formula for this slope difference as a function of $(a, \beta)$ 
and confirmed its validity. 

Regarding the observational feasibility of using $V(r)$ as a cosmological probe, the major obstacle is nothing but the noisy measurements of the peculiar velocities of real galaxies 
(e.g., see ref.~\cite{AE09} for a review on this issue).  Although successful reconstructions of the large-scale peculiar velocity field from the observational galaxy surveys have been 
reported in several literatures~\cite{cou-etal23,vee-etal23,c4,hof-etal24}, the velocity reconstruction itself required a priori assumption on the background cosmology. 
To test the CIDER models with bound-zone velocity slope in practice,  it is highly desirable to make a cosmology independent estimation of $V(r)$ around massive clusters. 

As shown in the previous works~\cite{fal-etal14,lee-etal15,bri-etal16,LY16,lee17,han-etal19}, it is possible to estimate the peculiar velocities of bound-zone galaxies around massive target clusters 
from observations with information only on their redshifts if the target clusters satisfy the following two conditions.  First, their masses have already been estimated with high accuracy. 
Second, the bound-zone galaxies are located in anisotropic web environments like filaments and sheets whose axes are not aligned with the line of sight directions~\cite{fal-etal14,bri-etal16}.  
As most of the galaxies in the universe are known to be embedded in filaments~\cite{web}, the second condition does not impose any severe observational limitation on the section of targets. 
Whereas, the first condition would restrict the target selection to those well relaxed low-$z$ clusters ($z\ll 1$) since the mass measurements of high-$z$ clusters in the middle of merging 
often suffer from large uncertainties~\cite{cl_mass}.
Nevertheless,  given that the upcoming next-generation weak lensing surveys like the Legacy Survey of Space and Time5~\cite{lsst} are expected to achieve the unprecedentedly precise 
measurements of the masses of high-$z$ clusters, we conclude that the bound-zone velocity slope holds a good future prospect as a powerful complimentary probes of those 
non-standard cosmologies whose expansion histories are the same as the $\Lambda$CDM model. 
 
 \acknowledgments

JL acknowledges the support by Basic Science Research Program through the NRF of Korea funded by the Ministry of Education (No.2019R1A2C1083855). 
MB acknowledges support by the project “Combining Cosmic Microwave Background and Large Scale Structure data: 
an Integrated Approach for Addressing Fun- damental Questions in Cosmology”, funded by the MIUR Progetti di Ricerca 
di Rilevante Interesse Nazionale (PRIN) Bando 2017 - grant 2017YJYZAH.

\end{document}